\documentclass[preprint,authoryear,12pt]{elsarticle}
\usepackage{graphicx}
\usepackage{amssymb,amsfonts,amsmath}
\usepackage{bm}

\usepackage{color}

\begin{document}

\newcommand{\EQ}{Eq.~}
\newcommand{\EQS}{Eqs.~}
\newcommand{\FIG}{Fig.~}
\newcommand{\FIGS}{Figs.~}

\begin{frontmatter}

\title{Evolution via imitation among like-minded individuals}

\author[utokyo]{Naoki Masuda\corref{cor1}
}
\address[utokyo]{Department of Mathematical Informatics,
The University of Tokyo,
7-3-1 Hongo, Bunkyo, Tokyo 113-8656, Japan}
\cortext[cor1]{masuda@mist.i.u-tokyo.ac.jp}

\end{frontmatter}

\section*{Abstract}

In social situations with which evolutionary game is concerned, individuals are considered to be heterogeneous in various aspects. In particular, they may differently perceive the same outcome of the game owing to heterogeneity in idiosyncratic preferences, fighting abilities, and positions in a social network. In such a population, an individual may imitate successful and similar others, where similarity refers to that in the idiosyncratic fitness function. I propose an evolutionary game model with two subpopulations on the basis of multipopulation replicator dynamics to describe such a situation. In the proposed model, pairs of players are involved in a two-person game as a well-mixed population, and imitation occurs within subpopulations in each of which players have the same payoff matrix. It is shown that the model does not allow any internal equilibrium such that the dynamics differs from that of other related models such as the bimatrix game. In particular, even a slight difference in the payoff matrix in the two subpopulations can make the opposite strategies to be stably selected in the two subpopulations in the snowdrift and coordination games.

\bigskip

Keywords: evolutionary game; replicator dynamics; homophily; cooperation

\newpage

\section{Introduction}\label{sec:introduction}

A basic assumption underlying many
evolutionary and economic game theoretical models
is that individuals are the same except for
possible differences in the strategy that they select. In fact,
a population of individuals involved in ecological or social interaction
is considered to be heterogeneous. For example, different individuals
may have different fighting abilities or endowments
\citep{Landau1951BullMathBiophys1,Hammerstein1981AnimBehav,Maynardsmith1982book,Mcnamara1999Nature}, 
occupy different positions in
contact networks specifying the peers with whom the game is played
\citep{Szabo2007PhysRep,Jackson2008book}, or
have different preferences over the objective outcome of the
game. The last situation is succinctly represented by the
Battle of the Sexes game in which
a wife and husband prefer to go to watch
opera and football, respectively, whereas their stronger priority is on going out
together \citep{Luce1957book} (the Battle of
the Sexes game here is different from the 
one that models conflicts between males
and females concerning parental investment as described in \cite{Dawkins1976book,Schuster1981AnimBehav,Maynardsmith1982book,Hofbauer1988book,Hofbauer1998book}). In behavioral game experiments, the heterogeneity of subjects is rather a norm than exceptions
(e.g., \cite{Camerer2003book}).
For example, some humans are cooperative in the public goods game and others are not (e.g., \cite{Fischbacher2001EconLett,Jacquet2012CommIntegBiol}), and some punish non-cooperators more than others do \citep{Fehr2002Nature,Dreber2008Nature}.

Evolution of strategies in such a heterogeneous population is the
focus of the present paper. This
question has been examined along several lines.

First, in theory of preference,
it is assumed that individuals maximize their own idiosyncratic
utilities that vary between individuals.
The utility generally deviates from the fitness
on which evolutionary pressure operates (e.g., \cite{Sandholm2001RevEconDyn,Dekel2007RevEconStud,Alger2012JTheorBiol,Grund2013SciRep}).

In fact, experimental evidence shows that
individuals tend to imitate behavior of similar others in the context of
diffusion of innovations \citep{Rodgers2003book} and
health behavior \citep{Centola2011Science}.
Also in the context of economic behavior described as games, individuals may preferentially imitate similar others because similar individuals are expected to be 
interested in maximizing
similar objective functions. This type of behavior is not considered in previous preference models in which
individuals can instantaneously maximize their own payoffs, and selection occurs on the basis of the fitness function common to the entire population.
The model proposed in this study deals with
evolutionary dynamics in which individuals in a heterogeneous population 
mimic successful
and similar others. The similarity here refers to that in the
idiosyncratic preference.

Second, evolution in heterogeneous populations has been investigated
with the use of the evolutionary bimatrix game
\citep{Hofbauer1988book,Hofbauer1998book,Weibull1995book}.
A payoff bimatrix describes the payoff imparted to the two players
in generally asymmetric roles. In its evolutionary dynamics,
a population is divided into two subpopulations,
pairs of individuals selected from the different subpopulations play the
game, and selection occurs within each subpopulation.
The population then has
bipartite structure induced by the fixed role of individuals.
However, the most generic population structure 
for investigating interplay of evolution via social learning and
idiosyncratic preferences would be a well-mixed
population without fixed roles of individuals.

Third, evolutionary game dynamics on heterogeneous social networks
\citep{Szabo2007PhysRep}
is related to evolution in heterogeneous populations.
In most of the studies on this topic,
the payoff to an individual per generation is defined as
the obtained payoff summed over all the neighboring individuals. Then,
cooperation in social dilemma games is enhanced
on heterogeneous networks
\citep{Santos2005PRL,Duran2005PhysicaD,Santos2006PNAS}.
In this framework, hubs (i.e., those with many neighbors) and
non-hubs are likely to gain different payoffs mainly because of their positions in the contact network.
In particular, if the payoff of a single game is assumed to be nonnegative,
hubs tend to earn more than non-hubs simply because hubs have
more neighbors than non-hubs by definition \citep{Masuda2007RoyalB}.
However, as long as the contact network is fixed, a non-hub player
will not gain a large payoff by
imitating the strategy of a successful hub neighbor.
The number of neighbors serves as the resource of
a player. Then, it may be more natural to assume that players
imitate successful others with a similar number of neighbors.

Motivated by these examples,
I examine evolutionary dynamics in which
a player would imitate successful others
having similar preferences or inhabiting similar environments.
I divide the players
into two subpopulations depending on the subjective perception of the result of the game;
one may like a certain outcome of the game, and another may not like the same outcome. Imitation is assumed to occur within each subpopulation.
However, the interaction occurs as a well-mixed population. I also assume that
all the individuals have the same ability, i.e., no player is more likely
to ``win'' the game than others.

\section{Model}

Consider a population comprising
two subpopulations of players such that the payoff matrix depends on the
subpopulation. The payoff is equivalent to the fitness in the present model.
I call the game the subjective payoff game.
Each player, independent of the subpopulation, selects either of the two strategies denoted by $A$ and $B$.
The case with a general number of strategies can be analogously formulated.
The subjective payoff game and its replicator dynamics described in the following are a special case of the multipopulation game proposed before
\citep{Taylor1979JAP,Schuster1981BC-III} (for slightly different variants, see \cite{Maynardsmith1982book,Hofbauer1988book,Weibull1995book}).

The population is infinite, well-mixed, and consists of a fraction $p$
($0<p<1$) of type
$X$ players and a fraction $1-p$ of type $Y$ players.
The subjective payoff matrices that an $X$ player and a $Y$ player perceive as
row player are defined by
\begin{equation}
\bordermatrix{
 & A & B \cr
A & a_X & b_X \cr
B & c_X & d_X \cr} \;
\mbox{ and }\;
\bordermatrix{
 & A & B \cr
A & a_Y & b_Y \cr
B & c_Y & d_Y \cr},
\label{eq:payoff}
\end{equation} 
respectively. It should be noted that the payoff that an $X$ player, for example, perceives
depends on the opponent's strategy (i.e., $A$ or $B$) but not on the
opponent's type (i.e., $X$ or $Y$). The use of the two payoff matrices represents different idiosyncrasies in preferences in the two subpopulations. Alternatively, the payoff matrix differs by subpopulations because $X$ and $Y$ players have different tendencies to transform the result of the one-shot game (i.e., one of the four consequences composed of a pair of $A$ and $B$)
into the fitness. For example, $X$ and $Y$ 
players may benefit the most from mutual $A$ and mutual $B$, respectively.

The fractions of $X$ and
$Y$ players that select strategy $A$ are denoted by $x$ and
$y$, respectively. The fractions of $X$ and $Y$ players that select
strategy $B$ are equal to $1-x$ and $1-y$, respectively.
The payoffs to an $X$ player with strategies $A$ and $B$ are given by
\begin{equation}
\pi_{X,A} = a_X\left[px+(1-p)y\right]+b_X\left[p(1-x)+(1-p)(1-y)\right]
\label{eq:pi_XA}
\end{equation}
and
\begin{equation}
\pi_{X,B} = c_X\left[px+(1-p)y\right]+d_X\left[p(1-x)+(1-p)(1-y)\right],
\label{eq:pi_XB}
\end{equation}
respectively. The payoff to a $Y$ player is defined with $X$ replaced by $Y$
in Eqs.~\eqref{eq:pi_XA} and \eqref{eq:pi_XB}.

I assume that in the evolutionary dynamics, the players can only
copy the strategies of peers in the same subpopulation. This assumption reflects the premise that
the payoff in the present model is subjective such that the only comparison
that makes sense is that between the players in the same subpopulation.
The replicator dynamics of the subjective payoff game is then defined by
\begin{align}
\dot{x} =& x\left[
  \pi_{X,A}-\left(x\pi_{X,A}+(1-x)\pi_{X,B}\right)\right]\notag\\
=& x(1-x)\left\{(a_X-c_X)\left[px+(1-p)y\right] +
  (b_X-d_X)\left[p(1-x)+(1-p)(1-y)\right]\right\}
\label{eq:dx/dt M=2}
\end{align}
and
\begin{align}
\dot{y} =& y(1-y)
\left\{(a_Y-c_Y)\left[px+(1-p)y\right] +
  (b_Y-d_Y)\left[p(1-x)+(1-p)(1-y)\right]\right\},
\label{eq:dy/dt M=2}
\end{align}
where $\dot{x}$ and $\dot{y}$ represent the time derivatives.

\section{General results}

\subsection{Absence of internal equilibrium}

If $(x,y)$ is an internal equilibrium (i.e., $0<x,
y<1$) of the replicator dynamics given by \EQS\eqref{eq:dx/dt M=2} and \eqref{eq:dy/dt M=2}, 
$(a_X-c_X)\left[px+(1-p)y\right] +
  (b_X-d_X)\left[p(1-x)+(1-p)(1-y)\right]=(a_Y-c_Y)\left[px+(1-p)y\right] +
  (b_Y-d_Y)\left[p(1-x)+(1-p)(1-y)\right]=0$ must be
    satisfied. However, this is impossible unless a degenerate condition
$(a_X-c_X)(b_Y-d_Y)=(a_Y-c_Y)(b_X-d_X)$ is
satisfied. Therefore, for a generic pair of payoff matrices, the
replicator dynamics does not have an
internal equilibrium. 

Three remarks are in order. First, the absence of internal equilibrium
implies that the present dynamics does not allow limit cycles
 \citep{Hofbauer1988book,Hofbauer1998book}. 
Second, the present result contrasts with that for 
a two-subpopulation dynamics in which the perceived payoff matrix depends on 
the opponent's subpopulation
as well as on the focal player's subpopulation. In the latter case, an internal equilibrium or limit cycle can exist \citep{Schuster1981BC-III}.
Third, the present conclusion is different from that for
the bimatrix game. In the bimatrix game, each player in subpopulation $X$
exclusively interacts with each player in subpopulation $Y$.
Then, an internal equilibrium can exist, whereas, when it exists,
it is either a saddle
or a neutrally stable point surrounded by periodic orbits
\citep{Pohley1979Bios,Selten1980JTB,Schuster1981BC-II,Maynardsmith1982book,Hofbauer1988book,Hofbauer1998book}.

\subsection{Invariance under the transformation of payoff matrices}

The replicator dynamics without population structure, which is
referred to as the ordinary replicator dynamics in the following, is invariant under some transformations of the payoff matrix.
The dynamics given by
\EQS~\eqref{eq:dx/dt M=2} and \eqref{eq:dy/dt M=2} is also invariant under
some payoff transformations.

First, trajectories of the ordinary replicator equation are invariant under the addition of a common constant to all the entries of the payoff matrix. Similarly,
replacing the payoff matrices given by \EQ\eqref{eq:payoff} by
\begin{equation}
\begin{pmatrix}
a_X+h_X&b_X+h_X\\ c_X+h_X&d_X+h_X
\end{pmatrix}\; \mbox{ and }\;
\begin{pmatrix}
a_Y+h_Y&b_Y+h_Y\\ c_Y+h_Y&d_Y+h_Y
\end{pmatrix},
\label{eq:payoff constant added}
\end{equation}
where $h_X$ and $h_Y$ are arbitrary constants,
does not alter the dynamics.

Second, trajectories of the ordinary replicator equation are invariant under multiplication of all the entries of the payoff matrix by a common positive constant. It only changes the time scale. In the present model,
replacing \EQ\eqref{eq:payoff} by
\begin{equation}
\begin{pmatrix}
ka_X&kb_X\\ kc_X&kd_X
\end{pmatrix}\; \mbox{ and }\;
\begin{pmatrix}
ka_Y&kb_Y\\ kc_Y&kd_Y
\end{pmatrix},
\label{eq:payoff multiplied by k}
\end{equation}
where $k>0$, does not alter the dynamics. It should be noted that the
multiplicative factor for the two payoff matrices has to be the same for the
dynamics to be conserved.

Third, in the ordinary replicator equation, adding a common constant to any column of the payoff matrix does not alter a trajectory. In the present model, replacing \EQ\eqref{eq:payoff} by
\begin{equation}
\begin{pmatrix}
a_X+h_{X,A}&b_X+h_{X,B}\\ c_X+h_{X,A}&d_X+h_{X,B}
\end{pmatrix}\;\mbox{ and }\;
\begin{pmatrix}
a_Y+h_{Y,A}&b_Y+h_{Y,B}\\ c_Y+h_{Y,A}&d_Y+h_{Y,B}
\end{pmatrix}
\label{eq:payoff constant added to columns}
\end{equation}
does not alter the trajectory for arbitrary $h_{X,A}$, $h_{X,B}$, $h_{Y,A}$, and $h_{Y,B}$. This invariance is a generalization of the first invariance. It is
also equivalent to the invariance relationship found for a more general
model \citep{Schuster1981AnimBehav}.

\subsection{Condition for ESS}

Let us calculate the conditions for the combination of pure strategies
in each subpopulation to be evolutionarily stable strategies (ESSs).
Some definitions of ESS for multipopulation games exist
\citep{Taylor1979JAP,Schuster1981BC-III,Cressman1992book,Cressman1996TPB,Cressman2001JTB}, and it seems that consensus on the definition of the ESS in the case of multiple subpopulations
has not been reached \citep{Weibull1995book}. Here I adhere
to the definition given in \cite{Taylor1979JAP}
(also see \cite{Schuster1981BC-III}),
which 
was proposed for general two-subpopulation games in which intra-subpopulation and
inter-subpopulation interactions yield a different payoff to a focal player.

I start with stating the definition of the ESS by obeying \cite{Taylor1979JAP}.
Consider a general two-subpopulation matrix game such that there are $m$ and $n$ pure strategies in subpopulations $X$ and $Y$, respectively. A mixed strategy in $X$ and $Y$ is parametrized as $\bm x = (x_1\; x_2\; \cdots\; x_m)^{\top}$
and $\bm y = (y_1\; y_2\; \cdots\; y_n)^{\top}$, where $x_i$ ($1\le i\le m$) and 
$y_i$ ($1\le i\le n$) are the probabilities that the mixed strategy in $X$ and $Y$ takes the $i$th strategy, respectively, $\sum_{i=1}^m x_i = \sum_{i=1}^n y_i = 1$, and $\top$ denotes the transposition. Consider a population of resident players taking strategies $\bm x$ and $\bm y$ in subpopulations $X$ and $Y$, respectively. I assume that the payoff to a player adopting mixed strategy $\bm x^{\prime}$ in subpopulation $X$ embedded in this resident population is given by
\begin{equation}
\bm x^{\prime\top}(R_{XX}\bm x + R_{XY}\bm y).
\end{equation}
Similarly, assume that the payoff to mixed strategy $\bm y^{\prime}$ in subpopulation $Y$ embedded in the same resident population is given by
\begin{equation}
\bm y^{\prime\top}(R_{YX}\bm x + R_{YY}\bm y).
\end{equation}
Strategy $(\bm x, \bm y)$ is ESS if for any $(\bm x^{\prime}, \bm y^{\prime})\neq (\bm x, \bm y)$,
\begin{equation}
\bm x^{\prime\top}(R_{XX}\bm x + R_{XY}\bm y) +
\bm y^{\prime\top}(R_{YX}\bm x + R_{YY}\bm y)
\le
\bm x^{\top}(R_{XX}\bm x + R_{XY}\bm y) +
\bm y^{\top}(R_{YX}\bm x + R_{YY}\bm y).
\label{eq:def ESS 1}
\end{equation}
When the equality holds in Eq.~\eqref{eq:def ESS 1}, it is also required that
an additional condition given by
\begin{equation}
\bm x^{\prime\top}(R_{XX}\bm x^{\prime} + R_{XY}\bm y^{\prime}) +
\bm y^{\prime\top}(R_{YX}\bm x^{\prime} + R_{YY}\bm y^{\prime})
<
\bm x^{\top}(R_{XX}\bm x^{\prime} + R_{XY}\bm y^{\prime}) +
\bm y^{\top}(R_{YX}\bm x^{\prime} + R_{YY}\bm y^{\prime})
\label{eq:def ESS 2}
\end{equation}
is satisfied.

In the case of the subjective payoff game, I obtain
$m=n=2$,
\begin{align}
R_{XX} =& p\begin{pmatrix}
a_X & b_X\\ c_X & d_X
\end{pmatrix},
\label{eq:R_XX}\\
R_{XY} =& (1-p)\begin{pmatrix}
a_X & b_X\\ c_X & d_X
\end{pmatrix},
\label{eq:R_XY}\\
R_{YX} =& p\begin{pmatrix}
a_Y & b_Y\\ c_Y & d_Y
\end{pmatrix},
\label{eq:R_YX}\\
R_{YY} =& (1-p)\begin{pmatrix}
a_Y & b_Y\\ c_Y & d_Y
\end{pmatrix},
\label{eq:R_YY}
\end{align}
 $x_1=x$, $x_2=1-x$, $y_1=y$, and $y_2=1-y$. Therefore, Eqs.~\eqref{eq:def ESS 1} and \eqref{eq:def ESS 2} are reduced to
\begin{align}
&\left[(x-x^{\prime})(a_X-c_X\; b_X-d_X) +
(y-y^{\prime})(a_Y-c_Y\; b_Y-d_Y)\right]\notag\\
&\cdot \left[p\begin{pmatrix}x\\ 1-x\end{pmatrix}
+(1-p)\begin{pmatrix}y\\ 1-y\end{pmatrix}\right]\ge 0
\label{eq:def ESS 1 subjective}
\end{align}
and
\begin{align}
&\left[(x-x^{\prime})(a_X-c_X\; b_X-d_X) +
(y-y^{\prime})(a_Y-c_Y\; b_Y-d_Y)\right]\notag\\
&\cdot \left[p\begin{pmatrix}x^{\prime}\\ 1-x^{\prime}\end{pmatrix}
+(1-p)\begin{pmatrix}y^{\prime}\\ 1-y^{\prime}\end{pmatrix}\right]> 0
\label{eq:def ESS 2 subjective}
\end{align}
respectively.

\subsection{Pure strategy ESSs}

In this section, let us identify the pure strategy ESSs of the subjective payoff game.
First, suppose that the population in which all
players in both subpopulations adopt strategy $A$ is evolutionarily stable.
By substituting $x=y=1$ in Eq.~\eqref{eq:def ESS 1 subjective}, I obtain
\begin{equation}
(1-x^{\prime})(a_X-c_X) + (1-y^{\prime})(a_Y-c_Y)\ge 0.
\label{eq:ESS pure A 1}
\end{equation}
Because Eq.~\eqref{eq:ESS pure A 1} must hold true for $0\le x^{\prime}<1$ and $y^{\prime}=1$, a necessary condition reads
$a_X\ge c_X$.
If $a_X\ge c_X$ is satisfied with equality, Eq.~\eqref{eq:def ESS 2 subjective} for the same $0\le x^{\prime}<1$ and $y^{\prime}=1$, i.e.,
\begin{equation}
(1-x^{\prime})(a_X-c_X\; b_X-d_X)
\left[p\begin{pmatrix}x^{\prime}\\ 1-x^{\prime}\end{pmatrix}
+(1-p)\begin{pmatrix}1\\ 0\end{pmatrix}\right]> 0
\label{eq:ESS pure A 3}
\end{equation}
must be satisfied.
By substituting $a_X=c_X$ in Eq.~\eqref{eq:ESS pure A 3}, I obtain
$b_X>d_X$. The necessary conditions obtained so far are summarized as
\begin{equation}
a_X > c_X \mbox{ or } (a_X=c_X \mbox{ and } b_X>d_X).
\label{eq:ESS pure A final 1}
\end{equation}
These conditions are the same as the ESS conditions for the structureless population.
By considering the mutant parametrized by 
$x^{\prime}=1$ and $0\le y^{\prime}<1$, I similarly obtain the necessary
conditions for subpopulation $Y$ as
\begin{equation}
a_Y > c_Y \mbox{ or } (a_Y=c_Y \mbox{ and } b_Y>d_Y).
\label{eq:ESS pure A final 2}
\end{equation}

On the other hand, if \EQS\eqref{eq:ESS pure A final 1} and
\eqref{eq:ESS pure A final 2} are satisfied, \EQS\eqref{eq:def ESS 1 subjective} and \eqref{eq:def ESS 2 subjective} are satisfied for any
$(x^{\prime}, y^{\prime})\neq (1,1)$. Therefore, 
\EQS\eqref{eq:ESS pure A final 1} and
\eqref{eq:ESS pure A final 2} provide the necessary and sufficient conditions for strategy $A$ to be evolutionarily stable.
In conclusion, $A$ is evolutionarily stable for the entire population if $A$ is evolutionarily stable in each subpopulation in the ordinary sense.
Similarly, $B$ is an ESS of the subjective payoff game if $B$ is evolutionarily stable in each subpopulation.

Next, assume that the population in which all the players in subpopulations $X$ and $Y$ adopt $A$ and $B$, respectively, is evolutionarily stable.
By substituting $x=1$ and $y=0$
in Eq.~\eqref{eq:def ESS 1 subjective}, I obtain
\begin{equation}
(1-x^{\prime}) \left[(a_X-c_X)p + (b_X-d_X)(1-p)\right]
+ y^{\prime} \left[(a_Y-c_Y)p + (b_Y-d_Y)(1-p)\right]\ge 0.
\label{eq:ESS pure lopsided 1}
\end{equation}
Because Eq.~\eqref{eq:ESS pure lopsided 1} must hold true for $0\le x^{\prime}<1$ and $y^{\prime}=0$, a necessary condition reads
\begin{equation}
a_Xp+b_X(1-p)\ge c_Xp+d_X(1-p).
\label{eq:ESS pure lopsided 3}
\end{equation}
If Eq.\eqref{eq:ESS pure lopsided 3} is satisfied with equality,
Eq.~\eqref{eq:def ESS 2 subjective} for the same $0\le x^{\prime}<1$ and $y^{\prime}=0$, i.e.,
\begin{equation}
(1-x^{\prime})(a_X-c_X\; b_X-d_X)
\left[p\begin{pmatrix}x^{\prime}\\ 1-x^{\prime}\end{pmatrix}
+(1-p)\begin{pmatrix}0\\ 1\end{pmatrix}\right]> 0,
\label{eq:ESS pure lopsided 4}
\end{equation}
must be satisfied. By substituting
$a_Xp+b_X(1-p) = c_Xp+d_X(1-p)$ in Eq.~\eqref{eq:ESS pure lopsided 4},
I obtain
\begin{equation}
b_X>d_X.
\label{eq:b_X > d_X}
\end{equation}
Similarly, by considering
the mutant parametrized by 
$x^{\prime}=1$ and $0\le y^{\prime}<1$, I obtain
\begin{equation}
a_Yp+b_Y(1-p)\le c_Yp+d_Y(1-p).
\label{eq:ESS pure lopsided 5}
\end{equation}
When Eq.~\eqref{eq:ESS pure lopsided 5} is satisfied with equality, 
Eq.~\eqref{eq:def ESS 2 subjective} and
$a_Yp+b_Y(1-p) = c_Yp+d_Y(1-p)$ lead to
\begin{equation}
a_Y<c_Y.
\label{eq:a_Y < c_Y}
\end{equation}
The population given by $(x,y)=(1,0)$ is an ESS if Eqs.~\eqref{eq:ESS pure lopsided 3} and \eqref{eq:ESS pure lopsided 5} are satisfied, 
Eq.~\eqref{eq:b_X > d_X} holds true when
Eq.~\eqref{eq:ESS pure lopsided 3} is satisfied with equality,
and Eq.~\eqref{eq:a_Y < c_Y} holds true when 
Eq.~\eqref{eq:ESS pure lopsided 5} is satisfied with equality.
It should be noted that the conditions given by Eqs.~\eqref{eq:ESS pure lopsided 3} and \eqref{eq:ESS pure lopsided 5} depend on $p$.

\subsection{Non-equivalence to the bimatrix game}\label{sub:not bimatrix}

In this section, I show that the replicator equation of the subjective payoff game cannot be mapped to the replicator equation of a bimatrix game. It should be noted that the following arguments can be readily generalized to the case of an arbitrary number of strategies. 

In the bimatrix game in the well-mixed population, we consider all possible pairs of a player in subpopulation $X$ and a player in subpopulation $Y$. The two selected players are involved in a two person game, which is generally asymmetric. The payoff bimatrix is given by
\begin{equation}
\bordermatrix{
 & A & B \cr
A & (\tilde{a}_X,\tilde{a}_Y) & (\tilde{b}_X,\tilde{c}_Y) \cr
B & (\tilde{c}_X,\tilde{b}_Y) & (\tilde{d}_X,\tilde{d}_Y) \cr},
\end{equation}
where the first and second elements in each entry of the bimatrix represent the payoffs imparted to an $X$ player and $Y$ player, respectively. The row and column players correspond to subpopulations $X$ and $Y$, respectively.
The payoff to an $X$ player with strategies $A$ and $B$ is equal to
$\tilde{a}_Xy+\tilde{b}_X(1-y)$ and $\tilde{c}_Xy+\tilde{d}_X(1-y)$, respectively. Then, the replicator equation for subpopulation $X$ is given by
\begin{equation}
\dot{x} = x(1-x)\left[(\tilde{a}_X-\tilde{c}_X)y+(\tilde{b}_X-\tilde{d}_X)(1-y)\right].
\label{eq:dx/dt bimatrix game}
\end{equation}
Similarly,
\begin{equation}
\dot{y} = y(1-y)\left[(\tilde{a}_Y-\tilde{c}_Y)x+(\tilde{b}_Y-\tilde{d}_Y)(1-x)\right].
\label{eq:dy/dt bimatrix game}
\end{equation}

If the dynamics given by
\EQS\eqref{eq:dx/dt bimatrix game} and \eqref{eq:dy/dt bimatrix game}
is equivalent to that for the subjective payoff game (\EQS\eqref{eq:dx/dt M=2} and \eqref{eq:dy/dt M=2}), the
comparison of \EQS\eqref{eq:dx/dt M=2} and \eqref{eq:dx/dt bimatrix game}
yields $p(a_X-b_X-c_X+d_X)=0$ because the right-hand side of
Eq.~\eqref{eq:dx/dt M=2} must be independent of $x$ except for the multiplication factor $x(1-x)$. Because $p=0$ implies a structureless population,
$a_X-b_X-c_X+d_X=0$ holds true. Under this condition,
\EQ\eqref{eq:dx/dt M=2} is reduced to $\dot{x}=x(1-x)(b_X-d_X)$.
Then, $\tilde{a}_X-\tilde{b}_X-\tilde{c}_X+\tilde{d}_X=0$ and
$\tilde{b}_X-\tilde{d}_X = b_X-d_X$ must hold true. 
Similarly, $a_Y-b_Y-c_Y+d_Y=\tilde{a}_Y-\tilde{b}_Y-\tilde{c}_Y+\tilde{d}_Y=0$ and $\tilde{b}_Y-\tilde{d}_Y = b_Y-d_Y$ must hold true. Except for this
degenerate case, the two dynamics are not mapped from one to the other.

\section{Examples}

\subsection{Snowdrift game}

Consider the snowdrift game, also called the chicken game or the hawk-dove game, which represents a social dilemma situation
\citep{Sugden1986book,Hauert2004Nature}. The snowdrift game in two subpopulations in which the payoff matrix for a player depends on the opponent's subpopulation as well as the focal player's subpopulation is analyzed in \cite{Auger2001MMMAS}.
In the case without population structure,
a standard payoff matrix for the snowdrift game is given by
\begin{equation}
\bordermatrix{
 & A & B \cr
A & \beta-\frac{1}{2} & \beta-1 \cr
B & \beta & 0 \cr},
\end{equation}
where $\beta>1$. Strategies $A$ and $B$ correspond to cooperation and defection, respectively. If the opponent cooperates, it is better to defect. Otherwise, it is better to cooperate. The mixed population with a fraction of $A$ players given by $x^*=(2\beta-2)/(2\beta-1)$ is the unique ESS.

Consider the case in which the $\beta$ value depends on the subpopulation. Denote by $\beta_X$ and $\beta_Y$ the subpopulation-dependent $\beta$ value such that
\begin{equation}
\begin{pmatrix}
a_X & b_X\\ c_X & d_X
\end{pmatrix}
= \begin{pmatrix}
\beta_X-\frac{1}{2} & \beta_X-1\\ \beta_X & 0
\end{pmatrix}
\label{eq:payoff SD X}
\end{equation}
and
\begin{equation}
\begin{pmatrix}
a_Y & b_Y\\ c_Y & d_Y
\end{pmatrix}
= \begin{pmatrix}
\beta_Y-\frac{1}{2} & \beta_Y - 1\\ \beta_Y & 0
\end{pmatrix}.
\label{eq:payoff SD Y}
\end{equation}
Without loss of generality, I assume that $\beta_X>\beta_Y>1$.

Equations~\eqref{eq:dx/dt M=2} and \eqref{eq:dy/dt M=2} read, respectively,
\begin{align}
\dot{x} =& x(1-x)\left\{\left(\beta_X-1\right)-\left(\beta_X-\frac{1}{2}\right)\left[px+(1-p)y\right] \right\},\\
\dot{y} =& y(1-y)\left\{\left(\beta_Y-1\right)-\left(\beta_Y-\frac{1}{2}\right)\left[px+(1-p)y\right] \right\}.
\end{align}
Therefore, I obtain
\begin{equation}
\begin{cases}
\dot{x}>0, \dot{y}>0 & \mbox{ if } px+(1-p)y<\frac{\beta_Y-1}{\beta_Y-1/2},\\
\dot{x}>0, \dot{y}<0 & \mbox{ if }
\frac{\beta_Y-1}{\beta_Y-1/2} < px+(1-p)y < \frac{\beta_X-1}{\beta_X-1/2},\\
\dot{x}<0, \dot{y}<0 & \mbox{ if } px+(1-p)y>\frac{\beta_X-1}{\beta_X-1/2}.
\end{cases}
\label{eq:snowdrift dotx doty}
\end{equation}
Equation~\eqref{eq:snowdrift dotx doty} implies that
the replicator dynamics has the unique stable equilibrium whose location depends on the $p$ value.
If $0<p<(\beta_Y-1)/(\beta_Y-1/2)$,
the stable equilibrium is located at
\begin{equation}
(x^*, y^*) = \left(1, -\frac{p}{1-p}+\frac{\beta_Y-1}{(1-p)(\beta_Y-1/2)}\right).
\label{eq:SD equilibrium case 1}
\end{equation}
An example of this case is shown in \FIG\ref{fig:snowdrift}(a). This equilibrium is an ESS, which is shown in Appendix with the use of the ESS criterion established in 
\cite{Taylor1979JAP}.
If $(\beta_Y-1)/(\beta_Y-1/2)\le p\le (\beta_X-1)/(\beta_X-1/2)$,
the stable equilibrium is located at
\begin{equation}
(x^*, y^*) =(1, 0).
\label{eq:SD equilibrium case 2}
\end{equation}
An example of this case is shown in \FIG\ref{fig:snowdrift}(b). In this case, the equilibrium is an ESS because \EQS\eqref{eq:ESS pure lopsided 3}, \eqref{eq:b_X > d_X}, \eqref{eq:ESS pure lopsided 5}, and \eqref{eq:a_Y < c_Y} are satisfied.
Finally, if $(\beta_X-1)/(\beta_X-1/2)<p<1$,
the stable equilibrium is located at
\begin{equation}
(x^*, y^*) = \left(\frac{\beta_X-1}{p(\beta_X-1/2)}, 0\right).
\label{eq:SD equilibrium case 3}
\end{equation}
An example of this case is shown in \FIG\ref{fig:snowdrift}(c). This equilibrium is also 
an ESS (see Appendix for the proof).
In all three cases, $X$, i.e., the subpopulation
with the larger $\beta$ value, has a larger fraction of $A$ players than $Y$ does. 

In particular, at least one subpopulation is monomorphic for any $p$; $A$ monopolizes subpopulation $X$, or $B$ monopolizes subpopulation $Y$.
Even a slight difference in the payoff matrix (i.e., $\beta$ value) in the two subpopulations yields polarization of the strategies.
Similar polarization also occurs in the
bimatrix snowdrift game. However, 
the internal equilibrium exists but is a saddle in the case of the bimatrix game
\citep{Hofbauer1998book}
%
% [p.122]
such that the mechanism is different from that for the subjective payoff snowdrift game.

\subsection{Coordination game}

Consider a coordination game in which the players in different subpopulations have different preference of one strategy over the other. Specifically, let us set the payoff matrix for subpopulations $X$ and $Y$ to
\begin{equation}
\begin{pmatrix}
a_X & b_X\\ c_X & d_X
\end{pmatrix}
= \begin{pmatrix}
1+\alpha_X& 0\\ 0& 1-\alpha_X
\end{pmatrix}
\label{eq:payoff coord X}
\end{equation}
and
\begin{equation}
\begin{pmatrix}
a_Y & b_Y\\ c_Y & d_Y
\end{pmatrix}
= \begin{pmatrix}
1+\alpha_Y& 0\\ 0& 1-\alpha_Y
\end{pmatrix},
\label{eq:payoff coord Y}
\end{equation}
respectively, where $-1<\alpha_Y<0<\alpha_X<1$. Players in
$X$ and $Y$ prefer strategies $A$ and $B$, respectively.
Equations~\eqref{eq:payoff coord X} and \eqref{eq:payoff coord Y} can be regarded as a
payoff bimatrix of the Battle of the Sexes game \citep{Luce1957book}.
However, in the bimatrix game, 
the population is composed of two subpopulations corresponding to the
roles in the game. In contrast, the players in the present game do
not have roles and interact in a well-mixed population.

Given \EQS\eqref{eq:payoff coord X} and \eqref{eq:payoff coord Y}, \EQS\eqref{eq:dx/dt M=2} and \eqref{eq:dy/dt M=2} read
\begin{align}
\dot{x} =& x(1-x)\left[p(2x-1)+(1-p)(2y-1)+\alpha_X \right],
\label{eq:dx/dt coordination game}\\
\dot{y} =& y(1-y)\left[p(2x-1)+(1-p)(2y-1)+\alpha_Y \right].
\label{eq:dy/dt coordination game}
\end{align}

Near $x=y=1$,
$p(2x-1)+(1-p)(2y-1)+\alpha_X>p(2x-1)+(1-p)(2y-1)+\alpha_Y>0$ is
satisfied. Therefore, $(x^*,y^*)=(1,1)$ is a stable equilibrium of the replicator equations of the subjective payoff coordination game. 
Because \EQS\eqref{eq:ESS pure A final 1} and \eqref{eq:ESS pure A final 2}
are satisfied with \EQS\eqref{eq:payoff coord X} and \eqref{eq:payoff coord Y},
$(x^*,y^*)=(1,1)$ is an ESS. Similarly,
because $p(2x-1)+(1-p)(2y-1)+\alpha_Y<p(2x-1)+(1-p)(2y-1)+\alpha_X<0$ near
$x=y=0$, $(x^*,y^*)=(0,0)$
is a stable equilibrium.
Because \EQS\eqref{eq:payoff coord X} and \eqref{eq:payoff coord Y} yield $b_X<d_X$ and $b_Y<d_Y$, $(x^*,y^*)=(0,0)$ is another ESS.
These results are
consistent with those for the coordination game without population structure; the two strategies are bistable.

\begin{figure}[h]
\begin{center}
\includegraphics[width=8cm]{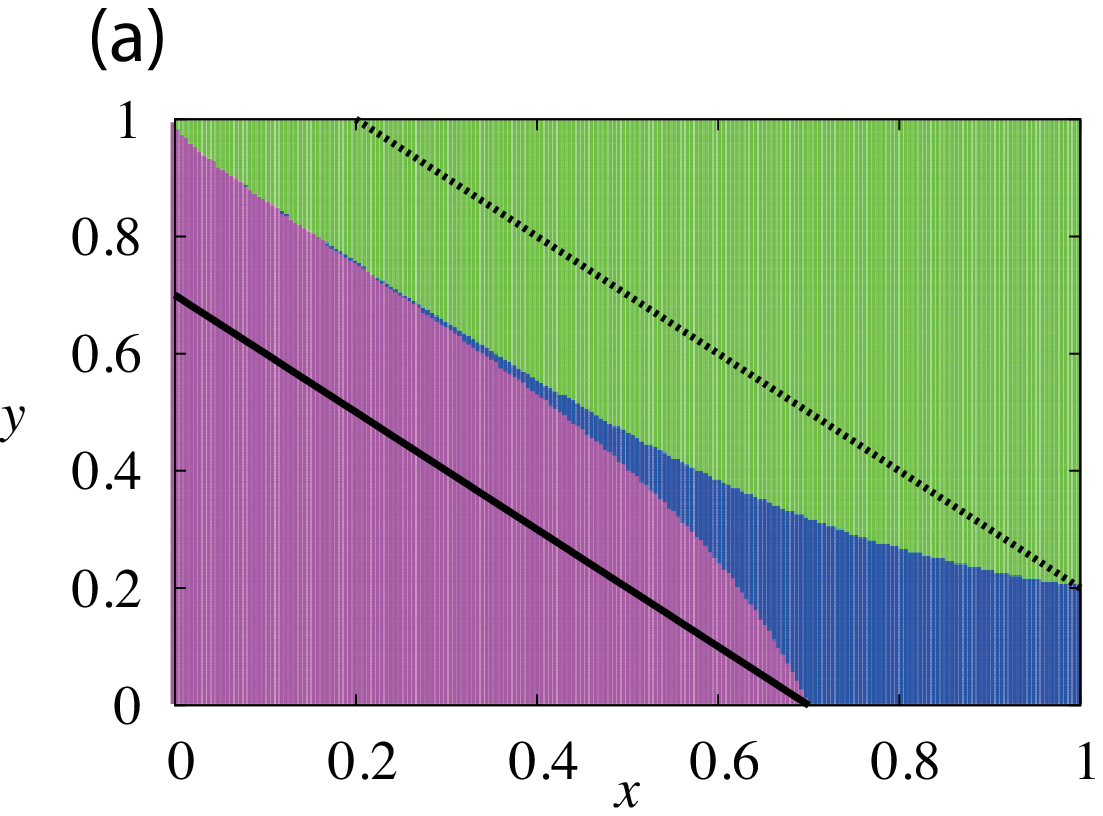}
\includegraphics[width=8cm]{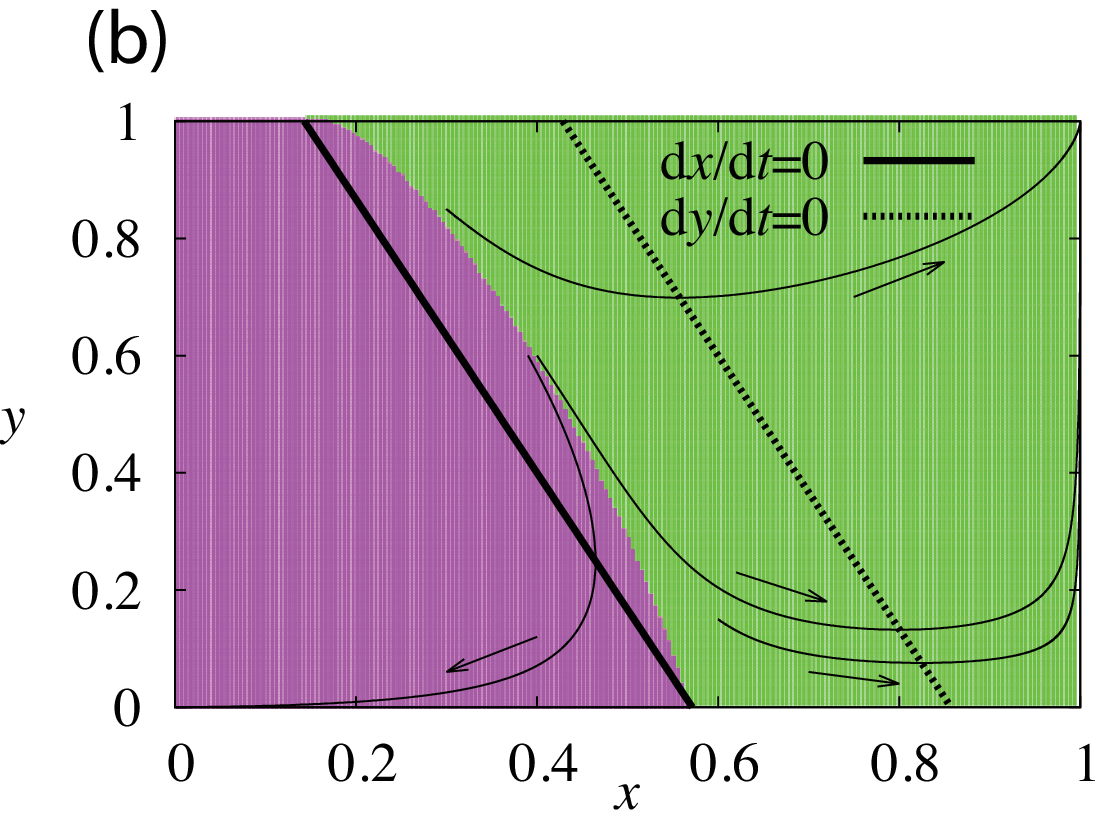}
\caption{Replicator dynamics of the subjective payoff coordination game. The magenta, green, and blue regions represent the attractive basins for $(x^*,y^*)=(0,0)$, $(1,1)$, and $(1,0)$, respectively.
The thick solid lines represent $\dot{x}=0$, i.e., 
$y=-px/(1-p)+(1-\alpha_X)/2(1-p)$. The thick dotted lines represent
$\dot{y}=0$, i.e., $y=-px/(1-p)+(1-\alpha_Y)/2(1-p)$.
(a) $p=0.5$, $\alpha_X=0.3$, and $\alpha_Y=-0.2$. 
(b) $p=0.7$, $\alpha_X=0.2$, and $\alpha_Y=-0.2$.
The thin solid curves in (b) represent trajectories converging to $(x^*,y^*)=(0,0)$ or $(1,1)$. For calculating the attractive basins and individual trajectories, the Euler scheme
with ${\rm d}t=0.005$ was used.}
\label{fig:coordination game}
\end{center}
\end{figure}

The subjective payoff version of the coordination game
yields two phenomena that are absent in the same game without population structure.
First, the polarized configuration in which all the $X$ players adopt $A$ and all the $Y$ players adopt $B$ is an stable equilibrium if
\begin{equation}
\frac{1-\alpha_X}{2}<p<\frac{1-\alpha_Y}{2}.
\label{eq:cnd for polarization}
\end{equation}
When \EQ\eqref{eq:cnd for polarization} is satisfied,
\EQS\eqref{eq:ESS pure lopsided 3} and \eqref{eq:ESS pure lopsided 5} are satisfied with inequality such that
this population is a pure-strategy ESS.
Equation~\eqref{eq:cnd for polarization} is satisfied if $p$ is close
to $1/2$ or if $\alpha_X$ or $-\alpha_Y (>0)$, i.e.,
the asymmetry in the liking of the two actions, is large. As an example, the attractive basins of the three equilibria for $p=0.5$,
$\alpha_X=0.3$, and $\alpha_Y=-0.2$ are shown in
\FIG\ref{fig:coordination game}(a). The final configuration
of the population depends on the initial condition.
It should be noted that stable coexistence of the opposite pure strategies does not occur in the bimatrix coordination game (i.e., Battle of the Sexes game).

Second, the fraction of players employing a strategy in a
subpopulation can non-monotonically change in time. Some non-monotonic
trajectories starting from different initial conditions are shown in
\FIG\ref{fig:coordination game}(b) with $p=0.7$, $\alpha_X=0.2$, and
$\alpha_Y=-0.2$.  $\dot{x}>0$ holds true to the right of the thick
solid line in \FIG\ref{fig:coordination game}(b). If the initial condition is located slightly right to this line,
$y$ first decreases because $\dot{y}<0$ holds true to the left of the thick dotted line. If $\dot{x} (>0)$ is not large, the trajectory eventually crosses the thick solid line ($\dot{x}=0$) such that $x$ starts to decrease. If the initial $\dot{x}$ value is large enough, the trajectory eventually crosses the thick dotted line ($\dot{y}=0$) such that $y$ starts to increase. In both cases, the trajectory shows non-monotonic behavior. Such a non-monotonic
behavior does not occur in the coordination game without population structure.

In extensions of the Ising model \citep{Galam1997PhysicaA} and the voter model
\citep{MasudaGibertRedner2010PRE,MasudaRedner2011JSM}, which are non-game population dynamics,
idiosyncratic preferences of individuals lead to coexistence of
different states, where states are equivalent to strategies in
games. The present results 
are consistent with these results
in that idiosyncratic preferences let multiple states coexist when unanimity necessarily occurs in the absence of idiosyncrasy.

\subsection{Iterated prisoner's dilemma}

In this section, I examine the possibility of cooperation in the
iterated prisoner's dilemma (IPD) in which the unconditional cooperation (C) and
unconditional defection (D) are strategies $A$ and $B$, respectively. 
I do not assume error
in action implementation and do
assume that a next round of the game between
the same pair of players occurs with probability $w$ ($0<w<1$).
The following results also hold true if C is replaced by
the tit-for-tat (TFT) or the so-called GRIM strategy.
TFT starts with cooperation and selects the previous action (i.e.,
cooperate or defect) selected by the opponent. 
GRIM strategy starts with
cooperation and switches to permanent defection once the opponent
ever defects. The invariance of the following results holds true because the payoff matrix for the IPD, i.e., Eq.~\eqref{eq:poff matrix IPD} below, does not change when
C is replaced by TFT or GRIM.

Consider a standard payoff matrix for the single-shot prisoner's dilemma given by
\begin{equation}
\bordermatrix{
 & {\rm C} & {\rm D} \cr
{\rm C} & b-c & -c \cr
{\rm D} & b & 0 \cr},
\end{equation}
where $b$ and $c$ are the benefit and the cost of cooperation and satisfy $b>c>0$.
The expected payoff matrix for the IPD in the structureless population is given by 
\begin{equation}
\bordermatrix{
 & {\rm C} & {\rm D} \cr
{\rm C} & \frac{b-c}{1-w} & -c \cr
{\rm D} & b & 0 \cr}.
\label{eq:poff matrix IPD}
\end{equation}
If
\begin{equation}
w>w_{\rm crit}\equiv \frac{c}{b},
\label{eq:w>c/b}
\end{equation}
the prisoner's dilemma is effectively transformed into a coordination game such that mutual cooperation by C and mutual defection by D are bistable \citep{Axelrod1984book,Nowak2006book}.

Let us consider the situation in which two subpopulations possess different
discount factors $w_X$ and $w_Y$. In other words, assume
\begin{equation}
\begin{pmatrix}
a_X & b_X\\ c_X & d_X
\end{pmatrix}
= \begin{pmatrix}
\frac{b-c}{1-w_X} & -c\\ b & 0
\end{pmatrix}
\label{eq:payoff PD X}
\end{equation}
and
\begin{equation}
\begin{pmatrix}
a_Y & b_Y\\ c_Y & d_Y
\end{pmatrix}
= \begin{pmatrix}
\frac{b-c}{1-w_Y} & -c\\ b & 0
\end{pmatrix}.
\label{eq:payoff PD Y}
\end{equation}
Because the duration of IPD is the same for the two players, I interpret that $X$ players put more emphasis on long-term benefits than $Y$ players. Specifically, I assume
\begin{equation}
w_Y<\frac{c}{b}<w_X.
\label{eq:cnd on w}
\end{equation}

Given \EQ\eqref{eq:cnd on w}, \EQ\eqref{eq:dy/dt M=2} implies
$\dot{y}<0$ for all $0\le x, y\le 1$. Therefore, D eventually occupies subpopulation $Y$. I examine the possibility that cooperation occurs in subpopulation $X$. On the line $y=0$, \EQ\eqref{eq:dx/dt M=2} is reduced to
\begin{equation}
\dot{x}=x(1-x)\left[\frac{(b-c)w_Xpx}{1-w_X}-c\right].
\label{eq:dx/dt IPD at y=0}
\end{equation}
Therefore, the population in which D dominates in both subpopulations, i.e., $(x^*,y^*)=(0,0)$, is always a stable equilibrium of the replicator equations of the subjective payoff IPD game. It is an ESS because \EQS\eqref{eq:payoff PD X} and \eqref{eq:payoff PD Y} imply $b_X<d_X$ and $b_Y<d_Y$, respectively.

Equation~\eqref{eq:dx/dt IPD at y=0} implies that the combination of C in subpopulation $X$ and D in subpopulation $Y$ is a stable equilibrium if
\begin{equation}
w_X > w_{X,{\rm crit}}\equiv \frac{1}{\left(\frac{b}{c}-1\right)p+1}.
\label{eq:w_x>w_xc}
\end{equation}
Because \EQ\eqref{eq:ESS pure lopsided 3} and
\EQ\eqref{eq:ESS pure lopsided 5} are satisfied with inequality
when \EQ\eqref{eq:w_x>w_xc} holds true,
this population is an ESS.
A large $p$ (i.e., large fraction of $X$ players) and a large
benefit-to-cost ratio $b/c$ lessen $w_{X,{\rm crit}}$ such that cooperation would occur.

\begin{figure}[h]
\begin{center}
\includegraphics[width=7cm]{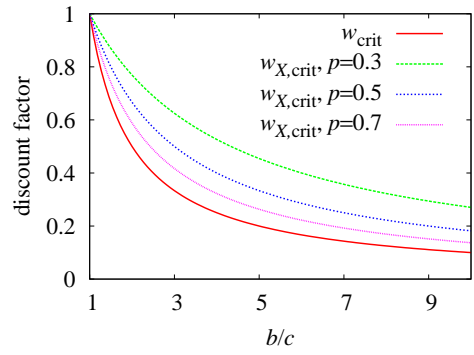}
\caption{Threshold discount factors above which C is locally stable in the IPD.
In the subjective payoff game, I set $p=0.3$, $p=0.5$, and $p=0.7$.}
\label{fig:IPD w_c}
\end{center}
\end{figure}

The threshold discount factors $w_{\rm crit}$ and $w_{X,{\rm crit}}$
are compared in
\FIG\ref{fig:IPD w_c} for some $p$ values.
The figure indicates that for a wide range of $p$, the condition for cooperation in the subjective payoff case is not very severe relative to the case without population structure.
In particular, both $w_{\rm crit}$ and $w_{X,{\rm crit}}$ tend to unity in the limit $b/c\to 1$. As $b/c\to\infty$, it follows that $w_{X,{\rm crit}}/w_{\rm crit}\to 1/p$ and both $w_{\rm crit}$ and $w_{X,{\rm crit}}$ converge to 0.
When $p=1/2$, condition~\eqref{eq:w_x>w_xc} coincides with the condition for the risk dominance of C over D in the structureless population, i.e., $w>2c/(b+c)$.

\section{Discussion}

I proposed the so-called subjective payoff game and its replicator dynamics.
The model is mathematically a special case of the previously analyzed
model with two subpopulations
\citep{Taylor1979JAP,Schuster1981BC-III}. However, the present model is motivated by the possibility that different players may perceive
the same result of the game to transform it to the fitness
in different manners. The model shows
polarization in the snowdrift and coordination games, non-monotonic time courses in the coordination game, and a wide margin of 
cooperation in the IPD.
Extension of the present model to the case of more than two strategies and more than two subpopulations is straightforward. Generalizing the present results for such extended models warrants future work.

The replicator dynamics of the subjective payoff game
is different from that of the bimatrix game
(section~\ref{sub:not bimatrix}). In addition, the subjective payoff game cannot be mapped to a model with strategy-dependent interaction rates, which does not have multiple subpopulations within each of which imitation occurs \citep{Taylor2006TPB}.
The subjective payoff game is also different from those in which interaction is confined in single subpopulations, such as group selection models \citep{Wilson1975PNAS,West2007JEB}, island model \citep{Taylor1992EvolEcol}, and evolutionary set theory \citep{Tarnita2009PNAS}.

The present model is distinct from previous models of evolution of preference
(e.g.,
\cite{Sandholm2001RevEconDyn,Dekel2007RevEconStud,Alger2012JTheorBiol,Grund2013SciRep})
and the so-called subjective game
\citep{Kalai1995GamesEconBehav,Matsushima1997JapEconRev,Oechssler2003GamesEconBehav}.
Both in these and present models, the preference, or the subjective
payoff, is assumed to be consistent within each individual
\citep{Gintis2009book_bounds}. In these previous models, the utility that a player
maximizes and the fitness on which the selection pressure operates are
different. A player in such a model is rational enough to be able to personally maximize the
player's idiosyncratic utility.  In the present model, as in
standard evolutionary models, a player is subjected to
bounded rationality and tends to imitate successful
others (i.e., social learning). The difference from standard evolutionary models is that, in the present model, each player limits the set of possible parents from whom the strategy is copied to those with the same idiosyncratic payoff. In this way, the player can pursue both maximization of fitness via social learning and consistency with the player's idiosyncratic preference.

The subjective payoff game does not allow internal equilibria regardless of the stability. This result has implications in games in which internal equilibria play an important role in structureless populations. In the snowdrift game, a mixture of the two strategies is stable under ordinary replicator dynamics. In contrast, in the subjective payoff game, a slight difference in the payoff matrices perceived by the two subpopulations leads to polarization such that the two subpopulations tend to select the opposite strategies.

The rock--scissors--paper game
comprises three strategies that cyclically dominate one another.
It is straightforward to show that there is no internal
equilibrium in the subjective payoff game with a general number
of strategies. Therefore,
the subjective payoff variant of the rock--scissors--paper
game lacks the internal equilibrium of any kind and limit cycles.
Such a game behaves very differently from
the same game played in the structureless population
\citep{Hofbauer1988book,Hofbauer1998book,Nowak2006book},
bimatrix population
\citep{Hofbauer1988book,Hofbauer1998book,Sato2002PNAS},
and two subpopulations with different social learning rates
\citep{Masuda2008JTB}; these models allow a unique internal equilibrium.
It may be interesting to examine the rock--scissors--paper
game under the current framework.

I assumed that players imitate others in the same subpopulation. In
fact, there may be competition of update rules between such players
and those that imitate from the entire population. Nevertheless, at
least near pure stable equilibria, the population is considered to be
resistant against invasion by mutants that imitate from the entire
population. To explain why, let us suppose that
$X$ and $Y$ players select $A$ and $B$ in the equilibrium. A mutant
that imitates from the entire population and attempts to invade
subpopulation $X$ would sometimes select $B$ because $Y$ players
select $B$. Because $A$, not $B$, is the best
response in this population, such a mutant is considered not able to invade
the subpopulation of resident players. Therefore, the imitation rule considered in the
present study is considered to have evolutionary stability, at least in this case.

\section*{Appendix}

In this section, I show that the equilibria given by
\EQS\eqref{eq:SD equilibrium case 1} and \eqref{eq:SD equilibrium case 3} are ESSs of the subjective payoff snowdrift game. To this end, I use a matrix criterion for the ESS \citep{Taylor1979JAP} accommodated to the case of the two-strategy game.

Assume that a population given by $(x^*, y^*)$ satisfies
\begin{align}
&\left(x^{\prime}\; 1-x^{\prime}\right)\left[R_{XX}\begin{pmatrix} x^*\\ 1-x^*\end{pmatrix}
+ R_{XY}\begin{pmatrix}y^*\\ 1-y^*\end{pmatrix} \right]\notag\\
\le&
\left(x^*\; 1-x^*\right)\left[R_{XX}\begin{pmatrix} x^*\\ 1-x^*\end{pmatrix}
 + R_{XY}\begin{pmatrix} y^*\\ 1-y^*\end{pmatrix}\right]
\label{eq:Taylor 1}
\end{align}
and
\begin{align}
&\left(y^{\prime}\; 1-y^{\prime}\right)\left[R_{YX}\begin{pmatrix}x^*\\ 1-x^*\end{pmatrix} + R_{YY}\begin{pmatrix}y^*\\ 1-y^*\end{pmatrix}\right]\notag\\
\le&
\left(y^*\; 1-y^*\right)\left[R_{YX}\begin{pmatrix}x^*\\ 1-x^*\end{pmatrix} + R_{YY}\begin{pmatrix}y^*\\ 1-y^*\end{pmatrix}\right]
\label{eq:Taylor 2}
\end{align}
for $(x^{\prime}, y^{\prime}) =$ $(0, 0)$, $(1, 0)$, $(0, 1)$, and $(1, 1)$. I also assume that Eqs.~\eqref{eq:Taylor 1} and \eqref{eq:Taylor 2} are satisfied with equality for $(x^{\prime}, y^{\prime}) = (0, 0)$ if $x^*<1$ and $y^*<1$, 
for $(x^{\prime}, y^{\prime}) = (1, 0)$ if $x^*>0$ and $y^*<1$, for $(x^{\prime}, y^{\prime}) = (0, 1)$ if $x^*<1$ and $y^*>0$, and for $(x^{\prime}, y^{\prime}) = (1, 1)$ if $x^*>0$ and $y^*>0$.
The criterion dictates that $(x^*, y^*)$ is an ESS if and only if
\begin{equation}
(x_1\; -x_1\; y_1\; -y_1)
\begin{pmatrix}
R_{XX} & R_{XY}\\ R_{YX} & R_{YY}
\end{pmatrix}
\begin{pmatrix}
x_1\\ -x_1\\ y_1\\ -y_1
\end{pmatrix}<0.
\label{eq:matrix criterion}
\end{equation}
Here, $x_1\neq 0$ ($y_1\neq 0$) if and only if the payoff of a pure $A$ player in subpopulation $X$ ($Y$) and that of a pure $B$ player in subpopulation $X$ ($Y$) are the same in the equilibrium of interest.

Under the snowdrift game, 
the assumptions for $(x^*, y^*)$ are satisfied when $(x^*, y^*)$ is given by
Eq.~\eqref{eq:SD equilibrium case 1} or \eqref{eq:SD equilibrium case 3}.
Substitution of \EQS\eqref{eq:R_XX}, \eqref{eq:R_XY}, \eqref{eq:R_YX}, \eqref{eq:R_YY}, \eqref{eq:payoff SD X}, and \eqref{eq:payoff SD Y} in \EQ\eqref{eq:matrix criterion} yields
\begin{equation}
\left[px_1+(1-p)y_1\right]\cdot \left[x_1\left(\frac{1}{2}-\beta_X\right)
+ y_1\left(\frac{1}{2}-\beta_Y\right)\right] < 0.
\label{eq:matrix criterion SD}
\end{equation}
For the equilibrium given by \EQ\eqref{eq:SD equilibrium case 1} to be
an ESS, \EQ\eqref{eq:matrix criterion SD} must be satisfied for
$x_1=0$ and $y_1\neq 0$.
For the equilibrium given by \EQ\eqref{eq:SD equilibrium case 3} to be
an ESS, \EQ\eqref{eq:matrix criterion SD} must be satisfied for
$x_1\neq 0$ and $y_1= 0$. In fact,
\EQ\eqref{eq:matrix criterion SD} is satisfied in both cases. Therefore, the two equilibria are ESSs.

\section*{Acknowledgements}

I thank Juli\'{a}n Garc\'{i}a and Arne Traulsen for valuable discussions and for reading of the manuscript. I also  
acknowledge the support provided through Grants-in-Aid for Scientific Research (No. 23681033) from MEXT, Japan, the Nakajima Foundation, and CREST, JST.

\newpage
\clearpage

\begin{figure}[h]
\begin{center}
\includegraphics[width=6cm]{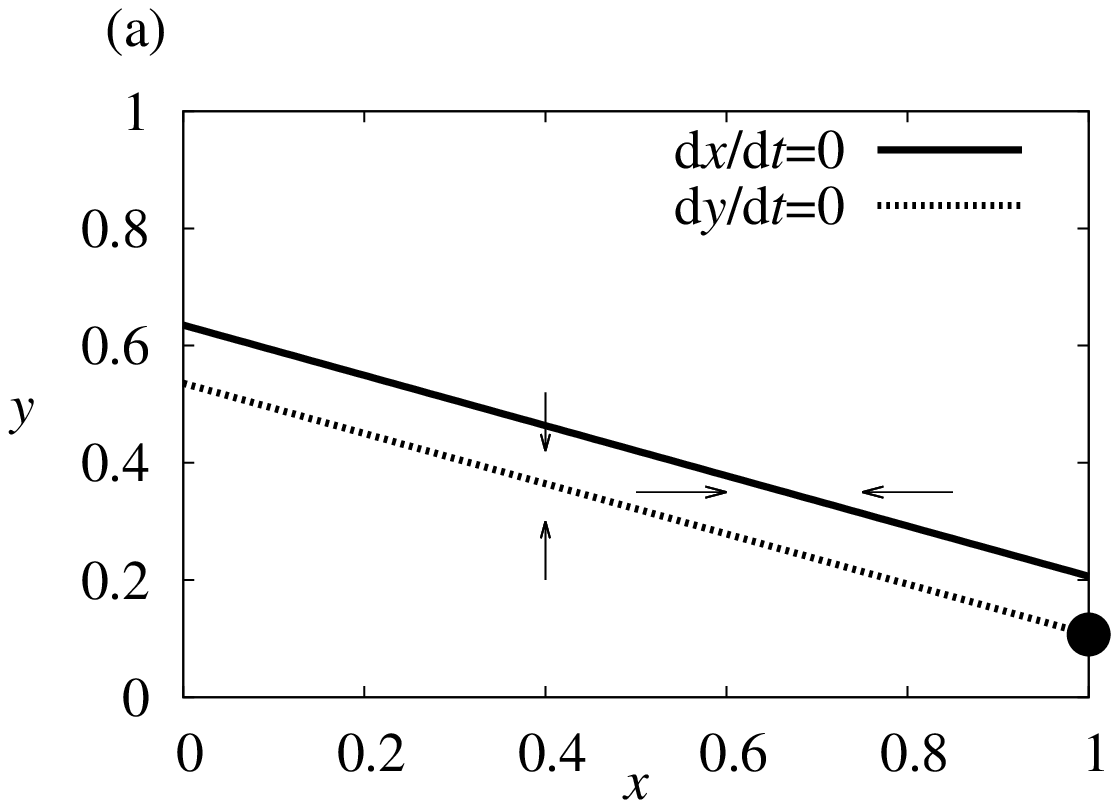}
\includegraphics[width=6cm]{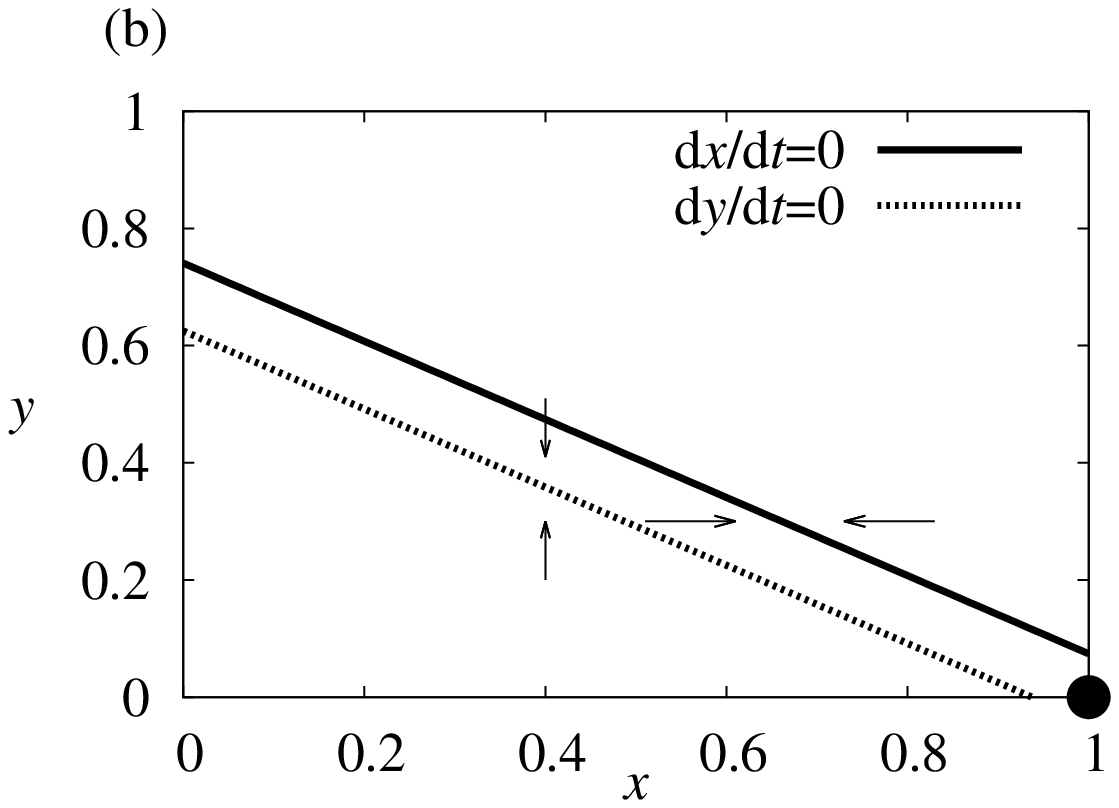}
\includegraphics[width=6cm]{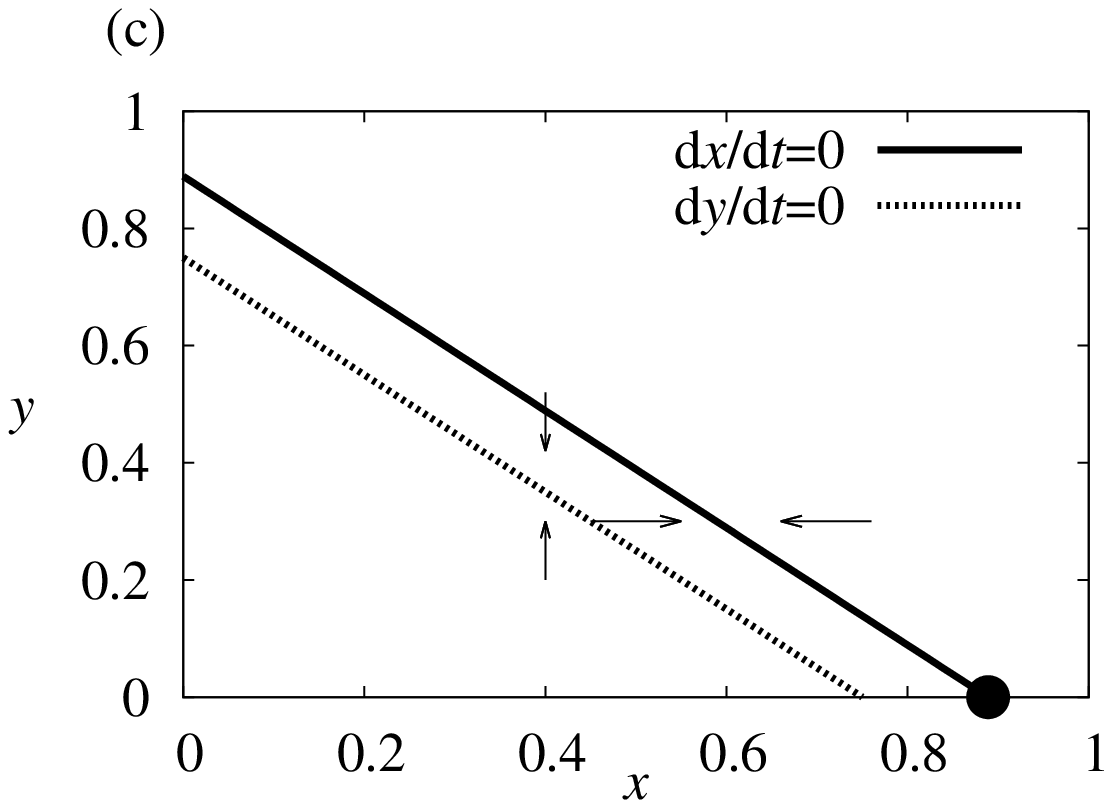}
\caption{The isoclines and direction field for the replicator dynamics of the subjective payoff snowdrift game with
$\beta_X=1.4$ and $\beta_Y=1.3$. The filled circles indicate the ESSs.
(a) $p=0.3$. (b) $p=0.4$. (c) $p=0.5$.}
\label{fig:snowdrift}
\end{center}
\end{figure}

\clearpage

\clearpage

\end{document}